# Emergence of the logarithmic average phonon frequency in the superconducting critical temperature formula

Nattawut Natkunlaphat, Pakin Tasee and Udomsilp Pinsook[*]

Department of Physics, Faculty of Science, Chulalongkorn University, Bangkok, 10330, Thailand

[*]Corresponding author's e-mail: Udomsilp.P@Chula.ac.th

**Abstract**

We analytically demonstrate the essential role of the logarithmic average phonon frequency, $\omega_{\ln}$, in describing the superconducting critical temperature, $T_c$, directly from the predictive function $\mathcal{L}(T_c; m_c) = \sum_{m=-m_c}^{m_c} \frac{\lambda_{m1}}{|2m-1|}$. The current study assumes that the Eliashberg spectral function, $\alpha^2 F(\omega)$, follows the Debye model in the low frequency spectrum ($0 \leq \omega \leq \omega_D$), whereas contributions from optical phonons dominate outside this range. Our findings confirm that, under the condition $1 < \frac{\omega_{\ln,D}}{2\pi T_c} \ll m_c$, $\mathcal{L}(T_c; m_c) \to \frac{\mathcal{L}(T_c)}{\lambda} \approx \ln\left(4e^\gamma \frac{\omega_{\ln}}{2\pi T_c}\right)$, resulting in the relation:

$$T_c^{NTP} \propto \omega_{\ln} \exp\left\{-\frac{1 + \mu^* \ln(4e^\gamma m_c)}{\lambda}\right\}.$$

Furthermore, comparisons of $T_c^{NTP}$ with the Allen-Dynes formula ($T_c^{AD}$), its modified version ($T_c^{AD(\text{Mod})}$), and the exact solutions ($T_c^{\text{Exact}}$) reveal notable correlations.

## 1. Introduction

For over 100 years, extensive research on superconductivity has been conducted. In 1957, the BCS theory [1] was proposed to explain phonon-mediated (conventional) superconductivity. One of the crucial results is the prediction of the superconducting critical temperature ($T_c$) formula in the weak electron-phonon (e-ph) coupling limit:

$$T_c^{BCS} \approx 1.13\omega_D \exp\left\{-\frac{1}{N(0)V}\right\}. \tag{1}$$

$T_c^{BCS}$ is scaled by the Debye frequency, $\omega_D$, where the exponential term involves the electronic density of state at the Fermi level, $N(0)$, and the pairing potential, $V$. Beyond the BCS framework, $T_c$ can be theoretically determined by solving the linearized Eliashberg gap equations (LEGE) [2–3]. Subsequently, in 1975, Allen and Dynes [4] developed the well-known empirical $T_c$ formula given by

$$T_c^{AD} = f_1 f_2 T_c^{\text{weak}}, \tag{2}$$

where

$$T_c^{\text{weak}} = \frac{\omega_{\ln}}{1.2} \exp\left\{-\frac{1.04(1+\lambda)}{\lambda - \mu^*(1+0.62\lambda)}\right\} \tag{3}$$

and

$$f_1 f_2 = \left[1 + \left(\frac{\lambda}{2.46(1+3.8\mu^*)}\right)^{3/2}\right]^{1/3}$$

$$\times \left[1 + \frac{\left(\frac{\omega_2}{\omega_{\ln}} - 1\right)\lambda^2}{\lambda^2 + \left[1.82(1+6.3\mu^*)\left(\frac{\omega_2}{\omega_{\ln}}\right)\right]^2}\right]. \tag{4}$$

The two correction factors $f_1$ and $f_2$ are essential for the strong e-ph coupling regime ($\lambda \gtrsim 1.5$). The earlier form of $T_c^{\text{weak}}$ proposed by McMillan [5] is proportional to $\omega_D$ rather than the logarithmic average phonon frequency, $\omega_{\ln}$. Another characteristic phonon frequency, $\omega_2$, naturally arises in the asymptotic behavior of $T_c$ [4, 6]:

$$T_c \to F(\mu^*)\sqrt{\lambda}\omega_2, \tag{5}$$

where $F(\mu^*)$ is a function of the Coulomb pseudopotential, $\mu^*$, stemming from the screened Coulomb interactions between electrons [7]. Typically, this parameter has a small value, ranging from 0 to 0.2. The larger value of $\mu^*$ is predicted as well in the Hubbard-Holstein model [8]. Based on the isotopic Eliashberg spectral function, $\alpha^2 F(\omega)$, we can define the following parameters:

$$\lambda \equiv \int_{\omega_{\min}}^{\omega_{\max}} \frac{2\alpha^2 F(\omega)}{\omega} \, d\omega, \tag{6}$$

$$\omega_{\ln} \equiv \exp\left\{\int_{\omega_{\min}}^{\omega_{\max}} g(\omega) \ln \omega \, d\omega\right\}, \tag{7}$$

and

$$\omega_2 \equiv \left[\int_{\omega_{\min}}^{\omega_{\max}} g(\omega)\omega^2 \, d\omega\right]^{1/2}. \tag{8}$$

Here, we introduce the normalized distribution function $g(\omega) \equiv \frac{2\alpha^2 F(\omega)}{\lambda \omega}$, where $\omega$ represents the phonon frequency. To be specific, there are two types of $\alpha^2 F(\omega)$: joint-shaped and disjoint-shaped. The joint-shaped $\alpha^2 F(\omega)$ does not exhibit a separation gap where $\alpha^2 F(\omega) = 0$ in the frequency domain $[\omega_{\min}, \omega_{\max}]$. This type of $\alpha^2 F(\omega)$ is commonly found in 3D materials, whereas the disjoint type is often observed in 2D materials. For practical purposes, we will primarily concentrate on the joint-shaped $\alpha^2 F(\omega)$.

In this decade, the search for superconducting materials with high $T_c$ continues to pose a challenge [9], particularly in the case of hydride compounds, as they are conventional superconductors [10–11]. In a publication [12], a support vector machine technique has been shown to exhibit significant accuracy in classifying metal polyhydrides with $T_c$ over 150 K, even without explicitly quoting the parameter space $\{\lambda, \omega_{\ln}, \omega_2, \mu^*\}$. Nevertheless, it is widely accepted that a set of parameters $\{\lambda, \omega_{\ln}, \omega_2, \mu^*\}$ is necessary for establishing (approximate) $T_c$ formulas. In addition to the Allen-Dynes perspective, this is evident in the machine-learning-driven $T_c$ formula [13] and the recent research conducted by Pinsook et al. [14]. An interesting issue is that the theoretical evidence supporting the essential role of $\omega_{\ln}$ in the $T_c$ equation remains ambiguous. Moreover, unlike $\lambda$ and $\omega_2$, $\omega_{\ln}$ cannot be directly measured from experiments. This motivates Pinsook et al. to propose an alternative solution to the LEGE based on the assumption of the largest first-order Matsubara gap function, resulting in the central inequality:

$$\mathcal{L}(T_c; m_c) \equiv \sum_{m=-m_c}^{m_c} \frac{\lambda_{m1}}{|2m-1|} \geq 1 + \lambda + \sum_{m=-m_c}^{m_c} \frac{\mu^*}{|2m-1|}. \tag{9}$$

The predictive function $\mathcal{L}(T_c; m_c)$ contains information about $T_c$ via the kernel:

$$\lambda_{mn} = \int_{\omega_{min}}^{\omega_{max}} \frac{2\omega\,\alpha^2 F(\omega)\,d\omega}{\omega^2 + 4\pi^2 T_c^2(m-n)^2}, \tag{10}$$

where the order $n = 1$ is taken into account. For convenience, we assume that the cutoff parameter $m_c$ is sufficiently large. Consequently,

$$\sum_{m=-m_c}^{m_c} \frac{1}{|2m-1|} \approx \ln(4e^\gamma m_c) + \frac{1}{2m_c} + \frac{1}{2m_c+1} \sim \ln(4e^\gamma m_c), \tag{11}$$

where $\gamma \approx 0.577$ is the Euler-Mascheroni constant. Indeed, $\sum_{m=-m_c}^{m_c} \frac{1}{|2m-1|}$ rapidly converges to $\ln(4e^\gamma m_c)$ even $m_c$ is not very large. In some limits, the function $\mathcal{L}$ is independent of $m_c$. Hence, it is possible to express $\frac{\mathcal{L}(T_c)}{\lambda} \equiv \frac{\mathcal{L}(a^2)}{\lambda}$, where the parameter $a \equiv \frac{\omega_{ln}}{2\pi T_c}$. By using the Einstein model: $\alpha^2 F^E(\omega) \propto \delta(\omega - \omega_E)$ [4] and the Debye model: $\alpha^2 F^D(\omega) \propto \omega^2$ [15], $\frac{\mathcal{L}(a^2)}{\lambda}$ can be approximately divided into two regimes depending on the value of $a$, instead of $\lambda$. First, for $a^2 < 0.5$, $\frac{\mathcal{L}(a^2)}{\lambda}$ is well described by a rational function of polynomials. Second, for $a^2 \geq 0.5$, $\frac{\mathcal{L}(a^2)}{\lambda}$ is dominated by the logarithmic behavior. The significant finding is that, for $a^2 \geq 0.5$, the leading-term approximation inequality for $T_c$ is given by

$$T_c \leq \frac{2e^{\gamma-1}}{\pi} \omega_{ln} \exp\left\{-\frac{1+\mu^*\ln(4e^\gamma m_c)}{\lambda}\right\}. \tag{12}$$

The equality in Eq. (12) serves as a good representative $T_c$ formula when compared with various experimental and DFT data. However, a theoretical study of $\frac{\mathcal{L}(a^2)}{\lambda}$ in general cases has not yet been conducted. In this work, we aim to rigorously demonstrate the relation $T_c \propto \omega_{ln}$ from $\frac{\mathcal{L}(a^2)}{\lambda}$ by extending the mathematical procedures outlined in [14]. In the next section, we discuss the results for the Einstein and Debye models. Subsequently, we derive the logarithmic behavior of the function $\mathcal{L}$ for a general $\alpha^2 F(\omega)$, which leads to Eq. (12). Finally, we highlight the effectiveness of Eq. (12) by comparing it with both the original and modified Allen-Dynes formulas, as well as with the exact solutions to Eq. (9).

## 2. Einstein and Debye models

Frequently, simplified models of $\alpha^2 F(\omega)$ are used to study several aspects of superconductivity. The simplest case is the Einstein model, which corresponds to an ideal system mediated by optical phonons with a single phonon frequency $\omega_E$, characterized by

$$\alpha^2 F^E(\omega) = \frac{\lambda \omega_E}{2} \delta(\omega - \omega_E). \tag{13}$$

This model offers valuable insights into the study of $T_c$ [4, 14] and electrical resistivity of a wide range of superconducting hydride compounds [16]. We can use Eqs. (6–8) to find

$$\lambda = \lambda, \omega_{\ln} = \omega_E, \text{ and } \omega_2 = \omega_E. \tag{14}$$

It is worth noting that the bandwidth (constant) model [10], defined as

$$\alpha^2 F^{BW}(\omega) = C > 0 \tag{15}$$

for $0 < \omega_{\min} \leq \omega \leq \omega_{\max}$, is regarded as the simplest generalization of the Einstein model [17]. In most materials, $\alpha^2 F(\omega)$ consists of both acoustic and optical regimes, with the acoustic regime typically appearing in the low-phonon-frequency range, where $\omega_{\min} = 0$. In theory, $\alpha^2 F(\omega)$ behaves as a monomial function, i.e., $\alpha^2 F(\omega) \propto \omega^n$ with $n \geq 1$, in the low-frequency limit ($\omega \to 0$), implying $\frac{\alpha^2 F(\omega)}{\omega} \propto \omega^{n-1}$. Such monomial models for $\alpha^2 F(\omega)$ have also been employed in studies of electron self-energy in certain superconducting materials [18–19]. The case $n = 2$ is known as the Debye model, representing the simplest form of $\alpha^2 F(\omega)$ for a crystalline solid mediated by acoustic phonons. It is explicitly given by

$$\alpha^2 F^D(\omega) = \frac{\lambda}{\omega_D^2} \omega^2, \tag{16}$$

where $0 \leq \omega \leq \omega_D$. By applying Eqs. (6)–(8), we obtain

$$\lambda = \lambda, \omega_{\ln} = \frac{\omega_D}{\sqrt{e}}, \text{ and } \omega_2 = \frac{\omega_D}{\sqrt{2}}. \tag{17}$$

However, the linear trend $\alpha^2 F(\omega) \propto \omega$ appears to better align with some amorphous solids [20].

At this stage, we provide a brief review on the calculation of the predictive function $\mathcal{L}$ using the Einstein model. By substituting Eq. (13) into Eq. (10), together with the condition $m_c \gg 1$, the left-hand side of Eq. (9) turns into

$$\frac{\mathcal{L}^E(a; m_c)}{\lambda} = 1 + a^2 \sum_{m=1}^{m_c} \frac{4m}{4m^2 - 1} \frac{1}{m^2 + a^2}. \tag{18}$$

Note that Eq. (18) is written in terms of the parameter $a \equiv \frac{\omega_E}{2\pi T_c}$. For this situation, $a = \frac{\omega_{\ln}}{2\pi T_c} = \frac{\omega_2}{2\pi T_c}$. In this context, we choose to denote $\frac{\mathcal{L}^E(a; m_c)}{\lambda}$ rather than $\frac{\mathcal{L}^E(a^2; m_c)}{\lambda}$. In the Appendix, we elucidate the structure of the uncertainty $\frac{\Delta \mathcal{L}^E(a; m_c)}{\lambda} \equiv \frac{[\mathcal{L}^E(a; m_c+1) - \mathcal{L}^E(a; m_c)]}{\lambda}$ in the limit $m_c \gg 1$, verifying that Eq. (18) remains valid as long as $\frac{\Delta \mathcal{L}^E(a; m_c)}{\lambda a^2} < O\left(\frac{1}{m_c^3}\right)$. As a starting point, we tentatively investigate the mathematical structure of $\frac{\mathcal{L}^E(a; m_c)}{\lambda}$ in two limits. First, for $a < 1$ (i.e., $a < m$ for all $m$), we then make use of the following expansion:

$$a^2 \sum_{m=1}^{m_c} \frac{4m}{4m^2 - 1} \frac{1}{m^2 + a^2} \approx \sum_{m=1}^{m_c} \frac{4m}{4m^2 - 1} \sum_{k=0}^{\infty} (-1)^k \left(\frac{a}{m}\right)^{2k+2}. \tag{19}$$

As addressed in [14], in the limit $a^2 \to 0$, Eq. (19) is simplified to

$$\frac{\mathcal{L}^E(a; m_c)}{\lambda} \approx 1 + 1.545 a^2 - \frac{1.390 a^4}{1 + a^2}. \tag{20}$$

Importantly, we observe that Eq. (20) provides a good approximation for $0 \le a \le 1$. Second, for $a > m_c \gg 1$, we similarly have

$$a^2 \sum_{m=1}^{m_c} \frac{4m}{4m^2 - 1} \frac{1}{m^2 + a^2} \approx \sum_{m=1}^{m_c} \frac{4m}{4m^2 - 1} \sum_{k=0}^{\infty} (-1)^k \left(\frac{m}{a}\right)^{2k}. \tag{21}$$

Under the extreme condition $a \gg m_c$, one readily sees that

$$\sum_{m=1}^{m_c} \frac{4m}{4m^2-1}\frac{a^2}{m^2+a^2} \approx \sum_{m=1}^{m_c} \frac{4m}{4m^2-1} \sim \ln(4e^{\gamma-1}m_c) + \frac{1}{2m_c}. \quad (22)$$

Eq. (22) indicates that $\frac{\mathcal{L}^E(a \gg m_c; m_c)}{\lambda}$ is independent of $a$. In other words, the information of $T_c$ cannot be deduced from $\mathcal{L}^E$. We speculate that restoring the dependence of $T_c$ on $\mathcal{L}^E$ requires the cutoff parameter $m_c$ to be related to $T_c$, as in the square-well model, where $\lambda_{mn} \approx \lambda$ [3, 14].

It should be emphasized that Pinsook et al. apply the results from Eq. (21) to derive the relation:

$$\frac{\mathcal{L}^E(a; m_c)}{\lambda} \approx \ln(4e^\gamma a) \quad (23)$$

under the conditions $a^2 \gg 1$ and $m_c^2 \gg a^2$, which contradict the necessary condition for the convergence of the series in Eq. (21), namely, $\left|\frac{m}{a}\right| < 1$; that is, $a > m_c$. In this paper, to explore the logarithmic form in Eq. (23), we examine the closed form of the summation in Eq. (18), given by

$$\frac{\mathcal{L}^E(a; m_c)}{\lambda} =$$

$$1 + \frac{4a^2}{1+4a^2}\left\{\begin{array}{l}\left[\psi\left(m_c+\frac{1}{2}\right)-\psi\left(\frac{1}{2}\right)-\frac{2m_c}{2m_c+1}\right]\\ +\text{Re }\psi(1+ia)-\text{Re }\psi(1+m_c+ia)\end{array}\right\}, \quad (24)$$

where Re $\psi(z)$ is the real part of the digamma function $\psi(z)$. To simplify Eq. (24), we invoke the condition $m_c \gg 1$ and expand the relevant functions as follows:

$$\psi\left(m_c + \frac{1}{2}\right) \approx \ln m_c + \frac{1}{24m_c^2} + O\left(\frac{1}{m_c^4}\right), \quad (25)$$

$$\text{Re }\psi(1+m_c+ia) \approx \ln m_c + \frac{1}{2m_c} + \frac{6a^2-1}{12m_c^2} + O\left(\frac{1}{m_c^4}\right), \quad (26)$$

and

$$\frac{2m_c}{2m_c + 1} \approx 1 - \frac{1}{2m_c} + \frac{1}{4m_c^2} + O\left(\frac{1}{m_c^3}\right). \quad (27)$$

As a result, $\frac{\mathcal{L}^E(a;m_c)}{\lambda}$ reduces to

$$1 + \frac{4a^2\{\ln(4e^{\gamma-1}) + \operatorname{Re}\psi(1+ia)\}}{1+4a^2} - \frac{1}{2}\left(\frac{a}{m_c}\right)^2, \quad (28)$$

where we use $\psi\left(\frac{1}{2}\right) = \ln(4e^\gamma)$. In essence, Eq. (28) becomes independent of the cutoff parameter $m_c$ in two scenarios. First, within the range $0 \leq a \lesssim m_c$, or more strictly $0 \leq a \ll m_c$, the term $\frac{1}{2}\left(\frac{a}{m_c}\right)^2$ is negligible. Second, in the limit $m_c \to \infty$ (boundary case), the term $\frac{1}{2}\left(\frac{a}{m_c}\right)^2 \to 0$. Therefore,

$$\frac{\mathcal{L}^E(a;m_c)}{\lambda} = 1 + \frac{4a^2}{1+4a^2}\{\ln(4e^{\gamma-1}) + \operatorname{Re}\psi(1+ia)\}. \quad (29)$$

In the former scenario, particularly for $0 \leq a \leq 1$, Eq. (29) is effectively approximated by Eq. (20). In the latter scenario, Eq. (29) can be viewed as the Einstein universal function, which asymptotically behaves as $\ln(4e^\gamma a)$ under the condition $1 \ll a < \infty$. One recognizes that $m_c \to \infty$ leads to the divergence of the summation in Eq. (11), which is accompanied by $\mu^*$. To prevent this, we rely on a finite $m_c$. Thus, we conclude that although $m_c$ is finite, it is possible to achieve $\frac{\mathcal{L}^E(a;m_c)}{\lambda} \equiv \frac{\mathcal{L}^E(a)}{\lambda} \approx \ln(4e^\gamma a)$ using the expansion of Eq. (29), provided $a \gg 1$ ($a > 1$ as a weaker constraint) and $a \ll m_c$. With the help of Eq. (9) and Eq. (11), the relation for $T_c$ is given by

$$T_c \leq \frac{2e^{\gamma-1}}{\pi}\omega_E \exp\left\{-\frac{1+\mu^*\ln(4e^\gamma m_c)}{\lambda}\right\}. \quad (30)$$

We proceed further by considering non-zero bandwidth spectral functions, initially focusing on the Debye model. The mathematical results derived earlier are applicable in this part. With the aid of Eq. (18), we rewrite $\mathcal{L}(T_c; m_c)$, as defined in Eq. (9), in the following form:

$$\frac{\mathcal{L}(T_c; m_c)}{\lambda} = 1 + \frac{2}{\lambda}\int_{\rho_{\min}}^{\rho_{\max}} d\rho\, \frac{\alpha^2 F(\omega)}{\rho}\left(\sum_{m=1}^{m_c} \frac{4m}{4m^2-1}\frac{\rho^2}{m^2+\rho^2}\right), \quad (31)$$

where $\rho \equiv \frac{\omega}{2\pi T_c}$. As discussed earlier, when $\rho \ll m_c$, i.e. $\frac{\omega_{max}}{2\pi T_c} \ll m_c$, the summation no longer significantly depends on $m_c$, leading to the following approximations: for $0 \leq \rho \leq 1$,

$$\sum_{m=1}^{m_c} \frac{4m}{4m^2-1} \frac{\rho^2}{m^2+\rho^2} \approx 1.545\rho^2 - \frac{1.390\rho^4}{1+\rho^2}, \tag{32}$$

and for $\rho \gg 1$,

$$\sum_{m=1}^{m_c} \frac{4m}{4m^2-1} \frac{\rho^2}{m^2+\rho^2} \approx \ln(4e^{\gamma-1}\rho). \tag{33}$$

To simplify the analysis, Eq. (33) is presumed to be satisfied for $1 < \rho \ll m_c$, because the summation rapidly converges to $\ln(4e^{\gamma-1}\rho)$ for $\rho > 1$. For this reason, the integral in Eq. (31) is the sum of two separate integrals:

$$\int_{\rho_{min}}^{1} d\rho\, \alpha^2 F(\omega) \left(1.545\rho - \frac{1.390\rho^3}{1+\rho^2}\right)$$

$$+ \int_{1}^{\rho_{max}} d\rho\, \alpha^2 F(\omega) \left(\frac{\ln(4e^{\gamma-1}\rho)}{\rho}\right). \tag{34}$$

Having established this, we are ready to employ $\alpha^2 F^D(\omega)$, Eq. (16), to examine Eq. (31). In this instance, we have $\rho_{min} = 0$ and $\rho_{max} = \frac{\omega_D}{2\pi T_c}$. Eq. (31) now reads

$$\frac{\mathcal{L}^D(T_c)}{\lambda} \approx 1 + 2\left(\frac{2\pi T_c}{\omega_D}\right)^2 \int_0^1 d\rho\, \rho^2 \left(1.545\rho - \frac{1.390\rho^3}{1+\rho^2}\right)$$

$$+ 2\left(\frac{2\pi T_c}{\omega_D}\right)^2 \int_1^{\frac{\omega_D}{2\pi T_c}} d\rho\, \rho^2 \left(\frac{\ln(4e^{\gamma-1}\rho)}{\rho}\right). \tag{35}$$

The consequence is

$$\frac{\mathcal{L}^D(T_c)}{\lambda} \approx 1 + 0.040\left(\frac{2\pi T_c}{\omega_D}\right)^2 + \left[\ln\left(4e^{\gamma-1}\frac{\omega_D}{2\pi T_c}\right) - \frac{1}{2}\right]. \tag{36}$$

We observe that by rewriting Eq. (36) in terms of $a \equiv \frac{\omega_{ln}}{2\pi T_c} = \frac{\omega_D}{2\pi T_c \sqrt{e}}$, notably yields

$$\frac{\mathcal{L}^D(T_c)}{\lambda} \approx \frac{0.04}{e}\frac{1}{a^2} + \ln(4e^\gamma a). \tag{37}$$

Clearly, the first term in Eq. (37) is negligible for $a > 1$. This indicates that $\frac{\mathcal{L}^D(T_c)}{\lambda}$ is predominantly governed by $\ln(4e^\gamma a)$ for $a > 1$. The outcome is in agreement with the Einstein model, except that the term $\ln(4e^\gamma a)$ is associated with $\omega_{\ln}$, instead of other characteristic phonon frequencies. Let us concisely elaborate on this issue. Suppose we wish to express Eq. (36) in terms of $b \equiv \frac{\omega_2}{2\pi T_c} = \frac{\omega_D}{2\pi T_c \sqrt{e}}$. The term inside the brackets becomes $\ln(4e^{\gamma-1}b) + \left(\ln\sqrt{2} - \frac{1}{2}\right)$. Finally, we arrive at the equation:

$$\frac{\mathcal{L}^D(T_c)}{\lambda} \approx \frac{0.02}{b^2} + \ln(4e^\gamma b) + \left(\ln\sqrt{2} - \frac{1}{2}\right), \tag{38}$$

which implies that $\frac{\mathcal{L}^D(b)}{\lambda} \approx \ln\left(4e^\gamma \sqrt{\frac{2}{e}} b\right)$ for $b > 1$. The only difference from $\frac{\mathcal{L}^D(a)}{\lambda}$ is due to the additional factor $\sqrt{\frac{2}{e}}$, which is nothing but the ratio $\frac{\omega_{\ln}}{\omega_2}$. If one insists on the form $\frac{\mathcal{L}^D(a)}{\lambda} \approx \ln(4e^\gamma a)$, the corresponding relation for $T_c$ is stated in Eq. (12). On the other hand, this is not a unique expression, as $T_c$ is also proportional to $\omega_2$, since the current model involves two independent parameters. That is to say, $T_c \propto \omega_D$ is a certain relationship. In a similar manner, the fitting equation for $T_c$ is put forward in [21]. Up to this point, we have analyzed $\frac{\mathcal{L}(T_c)}{\lambda}$ based on two simple phonon spectra, demonstrating consistent implications for $a > 1$, while maintaining $a \ll m_c$. Our next step is to investigate the universal form of $\frac{\mathcal{L}(T_c)}{\lambda} \approx \ln(4e^\gamma a)$ in a broader context.

### 3. Necessity of $\omega_{\ln}$ in the $T_c$ formula

We are motivated to explore the criteria under which the expression $\frac{\mathcal{L}(T_c)}{\lambda} \approx \ln(4e^\gamma a)$ applies to general $\alpha^2 F(\omega)$. At the outset, it is worthwhile to consider the more obvious case of the realistic optical phonon spectra, which are not localized at a single frequency like the Einstein spectrum. Certain types of compositions, such as boron-doped diamond [22], exhibit $\alpha^2 F(\omega)$ that differ from those of conventional materials, particularly due to having $\omega_{\min} \gg 0$.

According to Eq. (34), it is evident that if $\rho_{min} = \frac{\omega_{min}}{2\pi T_c} > 1$, then the first integral vanishes. Therefore, $\frac{\mathcal{L}(T_c)}{\lambda}$ becomes

$$\frac{\mathcal{L}(T_c)}{\lambda} \approx 1 + \frac{2}{\lambda} \int_{\rho_{min}}^{\rho_{max}} d\rho \, \alpha^2 F(\omega) \left( \frac{\ln(4e^{\gamma-1}\rho)}{\rho} \right), \tag{39}$$

which can be rearranged as follows:

$$\frac{\mathcal{L}(T_c)}{\lambda} \approx 1 + \int_{\omega_{min}}^{\omega_{max}} d\omega \, g(\omega) \ln\left( \frac{4e^{\gamma-1}}{2\pi T_c} \omega \right). \tag{40}$$

Surprisingly, Eq. (40) produces the expression:

$$\frac{\mathcal{L}(T_c)}{\lambda} \approx 1 + \ln\left( \frac{4e^{\gamma-1}}{2\pi T_c} \right) + \int_{\omega_{min}}^{\omega_{max}} d\omega \, g(\omega) \ln \omega = \ln(4e^\gamma a). \tag{41}$$

In the above derivation, we utilize the properties: $\int_{\omega_{min}}^{\omega_{max}} d\omega \, g(\omega) = 1$ and $\int_{\omega_{min}}^{\omega_{max}} d\omega \, g(\omega) \ln \omega = \ln \omega_{ln}$. A remark is that the condition $\frac{\omega_{min}}{2\pi T_c} > 1$ always guarantees $a = \frac{\omega_{ln}}{2\pi T_c} > 1$; however, the reverse argument does not necessarily hold. We expect that the logarithmic relationship $\frac{\mathcal{L}(T_c)}{\lambda} \propto \ln a$ is fundamentally caused by the contribution of optical phonons.

To clarify the last point we mentioned, we employ $\alpha^2 F(\omega)$ defined on the interval $0 \leq \omega \leq \omega_{max}$ such that

$$\alpha^2 F(\omega) = \begin{cases} \alpha^2 F^D(\omega), & 0 \leq \omega \leq \omega_D, \\ \alpha^2 F^{op}(\omega), & \omega_D < \omega \leq \omega_{max}. \end{cases} \tag{42}$$

Our approach is to determine the relative contributions of the acoustic and optical branches to the function $\frac{\mathcal{L}(T_c)}{\lambda}$. In the literature [23–24], a similar approach has been applied to examine $T_c$ for sulfur hydride, in which $\alpha^2 F(\omega)$ has a disjoint shape, and the separation gap between the acoustic and optical branches is at $\omega \approx 15$ THZ. The current principle assumes that in the low-frequency phonon modes, the phonon density of states $F(\omega)$ can be fitted to the Debye model: $F(\omega) \propto \omega^2$ for $0 \leq \omega \leq \omega_D$ [25]; therefore, $\alpha^2 F(\omega)$ is simplified to $\alpha^2 F^D(\omega)$. For $\omega_D < \omega \leq \omega_{max}$, $\alpha^2 F(\omega)$ is mainly contributed by optical phonons. To this end, we take a closer look at Eq. (34) and rearrange it as

$$\int_0^1 d\rho\, \alpha^2 F(\omega) \left(1.545\rho - \frac{1.390\rho^3}{1+\rho^2} - \frac{\ln(4e^{\gamma-1}\rho)}{\rho}\right)$$
$$+ \int_0^{\rho_{max}} d\rho\, \alpha^2 F(\omega) \left(\frac{\ln(4e^{\gamma-1}\rho)}{\rho}\right). \quad (43)$$

By substituting Eq. (43) into Eq. (31), we obtain the sum of two terms:

$$\frac{\mathcal{L}(T_c)}{\lambda} = \Lambda_1 + \Lambda_2, \quad (44)$$

where

$$\Lambda_1 \equiv \frac{1}{\lambda}\int_0^{\omega_0} d\omega\, \frac{2\alpha^2 F^D(\omega)}{\omega}\left(1.545\rho^2 - \frac{1.390\rho^4}{1+\rho^2} - \ln(4e^{\gamma-1}\rho)\right), \quad (45)$$

and

$$\Lambda_2 \equiv 1 + \frac{1}{\lambda}\int_0^{\omega_{max}} d\omega\, \frac{2\alpha^2 F(\omega)}{\omega}\ln(4e^{\gamma-1}\rho). \quad (46)$$

We introduce an additional constraint: $\Omega \cong 2\pi T_c$ must not exceed $\omega_{\ln,D} \equiv \frac{\omega_D}{\sqrt{e}}$.

To evaluate Eq. (45), we recall Eq. (16) to write

$$\frac{2\alpha^2 F^D(\omega)}{\omega} = \left(\frac{2\lambda_D}{\omega_D^2}\right)\omega, \quad (47)$$

where $\lambda_D$ represents the e-ph coupling constant due to $\alpha^2 F^D(\omega)$, giving rise to

$$\Lambda_1 \equiv \frac{\lambda_D}{\lambda}\left(\frac{2\pi T_c}{\omega_D}\right)^2 \int_0^1 d\rho\, 2\rho\left(1.545\rho^2 - \frac{1.390\rho^4}{1+\rho^2} - \ln(4e^{\gamma-1}\rho)\right). \quad (48)$$

The integral in Eq. (48) is approximately equal to 0.040, which is identical to the prefactor of $\left(\frac{2\pi T_c}{\omega_D}\right)^2$ in Eq. (36). Fortunately, the exact form of Eq. (46) was already specified in Eq. (41); that is, $\Lambda_2 \approx \ln\left(4e^{\gamma}\frac{\omega_{\ln}}{2\pi T_c}\right)$. Here, we denote

$$\lambda = \lambda_D + \lambda_{op}, \quad (49)$$

and

$$\lambda \ln \omega_{\ln} = \lambda_D \ln \omega_{\ln,D} + \lambda_{op} \ln \omega_{\ln,op}. \tag{50}$$

The quantities related to the optical branch are defined as follows:

$$\lambda_{op} \equiv \int_{\omega_D}^{\omega_{\max}} \frac{2\alpha^2 F^{op}(\omega)}{\omega} d\omega, \tag{51}$$

and

$$\omega_{\ln,op} \equiv \exp\left\{\int_{\omega_D}^{\omega_{\max}} \frac{2\alpha^2 F^{op}(\omega)}{\lambda_{op}\omega} \ln \omega \, d\omega\right\}. \tag{52}$$

Thus, we find the elegant result:

$$\frac{\mathcal{L}(T_c)}{\lambda} \approx \left(\frac{0.040}{e}\right) \frac{\lambda_D}{\lambda} \frac{1}{a_D^2} + \ln(4e^\gamma a), \tag{53}$$

where, as previously defined, $a = \frac{\omega_{\ln}}{2\pi T_c}$.

The key point is that $a_D$ and $a$ share a deep connection. From Eq. (50), it is apparent that

$$\omega_{\ln} = (\omega_{\ln,D})^{\frac{\lambda_D}{\lambda}} (\omega_{\ln,op})^{\frac{\lambda_{op}}{\lambda}}, \tag{54}$$

which implies the useful equation:

$$a = (a_D)^{\frac{\lambda_D}{\lambda}} (a_{op})^{\frac{\lambda_{op}}{\lambda}}. \tag{55}$$

One can see that both $\frac{\lambda_D}{\lambda}$ and $\frac{\lambda_{op}}{\lambda}$ are less than unity. Since $\omega_{\ln,D} < \omega_{\ln,op}$, it follows that $a_D < a_{op}$. In view of the constraint $\Omega \cong 2\pi T_c < \omega_{\ln,D}$, we immediately have $a_D > 1$, which also imposes the restriction $a > 1$. Based on these arguments, it is reasonable to omit the first term in Eq. (53), leaving

$$\frac{\mathcal{L}(T_c)}{\lambda} \approx \ln(4e^\gamma a). \tag{56}$$

One can observe that the condition $a_D > 1$ implies $\frac{\omega_D}{2\pi T_c} > \sqrt{e} > 1$, which is analogous to the condition $\frac{\omega_{\min}}{2\pi T_c} > 1$, as used in Eq. (39), reflecting that Eq. (56) is essentially influenced by the presence of the optical branch. As previously pointed out, if the optical branch is absent, Eq. (56) is still satisfied, i.e.,

$\frac{\mathcal{L}(T_c)}{\lambda} \to \frac{\mathcal{L}^D(T_c)}{\lambda}$, but the deduced relation $T_c \propto \omega_{\ln}$ is not definitive. Finally, the equation for $T_c$ derived from Eq. (56) is given by Eq. (12), which can equivalently be written as

$$T_c = \frac{2e^{\gamma-1}}{\pi}(\omega_{\ln,D})^{\frac{\lambda_D}{\lambda}}(\omega_{\ln,op})^{\frac{\lambda_{op}}{\lambda}}\exp\left\{-\frac{1+\mu^*\ln(4e^\gamma m_c)}{\lambda}\right\}. \tag{57}$$

As discussed in references [23–24], the low-$T_c$ phase, i.e., $T_c < \widetilde{\omega}_{ac} \ll \widetilde{\omega}_{op}$, takes the form:

$$T_c = constant \times (\widetilde{\omega}_{ac})^{\frac{\lambda_{ac}}{\lambda}}(\widetilde{\omega}_{op})^{\frac{\lambda_{op}}{\lambda}}\exp\left\{-\frac{1+\lambda}{\lambda-\mu^*}\right\}, \tag{58}$$

where $\widetilde{\omega}_{ac}$ and $\widetilde{\omega}_{op}$ are the characteristic phonon frequencies associated with the acoustic and optical branches, respectively. This expression is analogous to Eq. (57) in the case of $\mu^* = 0$. At this stage, we have successfully demonstrated the origin of $\omega_{\ln}$ in the $T_c$ equation directly from theory. To broaden our discussion, we argue that the derivation in this section is also applicable to the disjoint type of $\alpha^2 F(\omega)$. This can be achieved by redefining Eq. (42) in such a way that $\alpha^2 F^{op}(\omega)$ is located on the interval $\Omega < \omega \leq \omega_{max}$, where $\Omega > \omega_D$. By convention, we refer only to the equality in Eq. (12) as the analytic $T_c$ formula, denoted by $T_c^{NTP}$. Indeed, it serves as an upper bound for $T_c$ as a consequence of assuming the largest first-order Matsubara gap function. We subsequently evaluate the applicability of $T_c^{NTP}$ in the next section.

## 4. Testing the analytic $T_c$ formula

Up to this point, we have analyzed various mathematical aspects. In this section, we aim to compare the analytic result with $T_c$ calculated using several methods. Figure 1 illustrates the strong correlation between $T_c^{NTP}$ (with the cutoff parameter set to $m_c = 95$ and $m_c = 150$) and the values calculated by means of the Allen-Dynes formula, $T_c^{AD}$. The parameters $\lambda$, $\omega_{\ln}$, $\omega_2$, and $\mu^*$ are extracted from references [22, 27]. The value $m_c = 95$ was initially adopted by Pinsook et al. We observe that most data points are situated below or on the unity line, indicating that $T_c^{NTP} \geq T_c^{AD}$. Furthermore, the modified version, $T_c^{AD(\text{Mod})}$, in which the correction factor $f_2$ is slightly adjusted as

$$f_2 \to f_2^{\text{Mod}} = 1 + \frac{\left(\frac{\omega_2}{\omega_{\ln}} - 1\right) \lambda^2}{\lambda^2 + [1.82(1 + 6.3\mu^*)]^2}, \quad (59)$$

promotes $T_c^{AD(\text{Mod})} \geq T_c^{AD}$ (with equality in the Einstein model) [14, 26], thereby leading to a modest enhancement of the correlation within a specific range. The deviations from the unity line become substantial when $T_c^{NTP} \gtrsim 120$ K, because in the high-$T_c$ regime, the contribution of $\omega_2$ plays a vital role in both $T_c^{AD}$ and $T_c^{AD(\text{Mod})}$. This confirms the recent evidence presented by Pinsook et al., which indicates that in the low-$T_c$ regime, $T_c^{NTP}$ is nearly comparable to $T_c^{AD(\text{Mod})}$, while the former remains appreciably simpler in form.

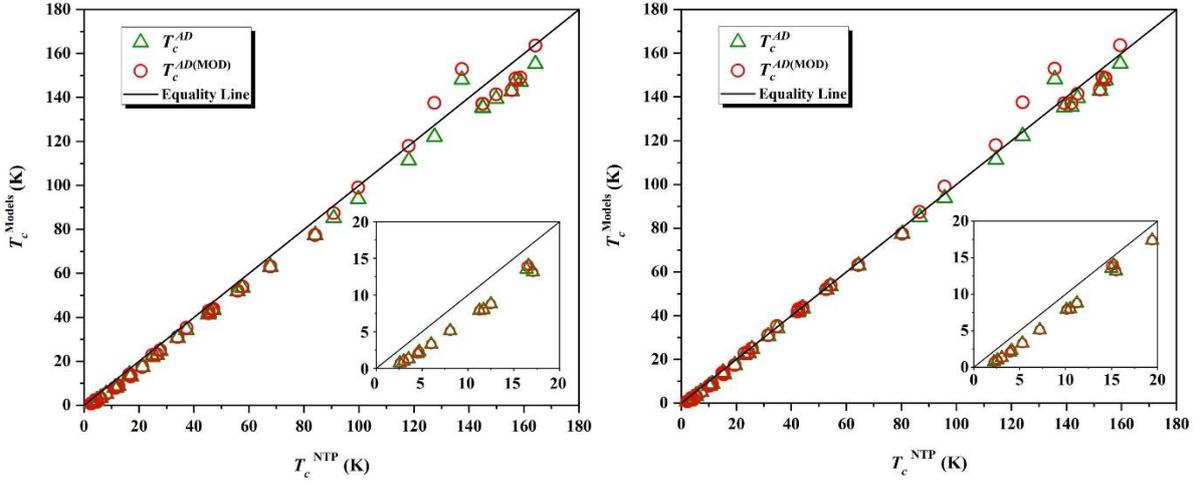

**Figure 1.** Plots of the numerical values calculated from the models: the Allen-Dynes formula, $T_c^{AD}$ (green triangles), and the modified Allen-Dynes formula, $T_c^{AD(\text{Mod})}$ (red circles), extracted from references [22, 27], against the analytic formula, $T_c^{NTP}$. The left panel ($m_c = 95$) and the right panel ($m_c = 150$) show a strong correlation between $T_c^{NTP}$ and the model-derived values, with $T_c^{AD(\text{Mod})}$ providing a marginally better agreement. The inset highlights the low-$T_c^{NTP}$ region, illustrating its higher values compared to those from the models. The effect of increasing $m_c$ results in a downward shift of data points in the right panel.

To justify the analytic formula within our framework, it is prudent to compare $T_c^{NTP}$ with the exact solution, $T_c^{\text{Exact}}$, obtained by numerically solving Eq. (9), as shown in figure 2. For this purpose, we employ the Eliashberg spectral functions reported in references [28–44]. We observe that, on average, the data points are distributed along the equality line; however, a few data points display discrepancies. It is worth noticing that the data points above the equality line, enclosed by the green circle, all share the characteristic $a_D = \frac{\omega_D}{2\pi\sqrt{e}T_c^{\text{Exact}}} < 1$ (assuming the lowest acoustic spectrum is fitted to the Debye model), whereas $a > 1$. In the high-$T_c^{NTP}$ region, we detect that three data points are clearly below the equality line, even though they correspond to values of $1 < a = \frac{\omega_{\ln}}{2\pi\sqrt{e}T_c^{\text{Exact}}} \lesssim 1.5$. A possible explanation for this is that they all possess the condition $a_D < 1$.

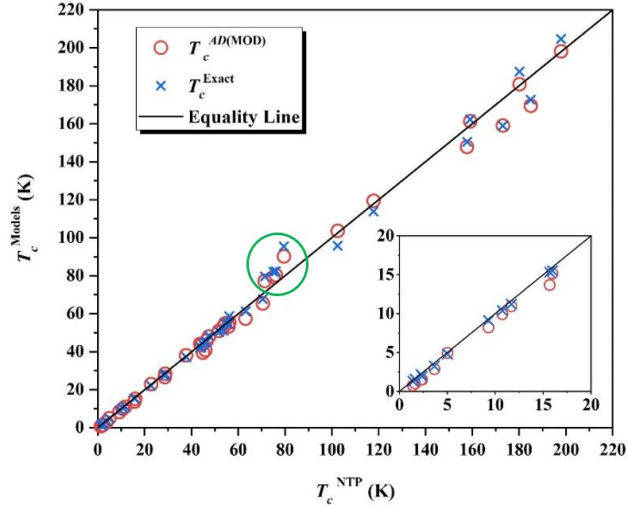

**Figure 2.** Plots of the numerical values calculated from the models: the exact numerical solutions to equation (9) with $m_c = 95$ using the Eliashberg spectral functions reported in references [28–44] (blue crosses) and the modified Allen-Dynes formula, $T_c^{AD(\text{Mod})}$ (red circles) in comparison with the analytic formula, $T_c^{NTP}$. An excellent agreement between $T_c^{NTP}$ and the outputs from the models is found, particularly in the low-$T_c^{NTP}$ region, as highlighted in the inset. The concerning discrepancies above the equality line, which correspond to $a_D < 1$, are marked by a green circle.

## 5. Conclusion

We analytically examine the predictive function $\frac{\mathcal{L}(T_c; m_c)}{\lambda}$ utilizing the Einstein and Debye models as guidance. Under the condition $\frac{\omega_{\max}}{2\pi T_c} \ll m_c$, the function $\frac{\mathcal{L}(T_c; m_c)}{\lambda}$ can be regarded as independent of $m_c$, and is thus denoted by $\frac{\mathcal{L}(T_c)}{\lambda}$. For the Einstein model, $\frac{\mathcal{L}^E(T_c)}{\lambda} \approx \ln(4e^\gamma a)$ if $a = \frac{\omega_E}{2\pi T_c} \gg 1$ ($a > 1$ as a weaker constraint), leading to $T_c \propto \omega_E$. By considering the Debye model, we recognize that the expression $\frac{\mathcal{L}^D(T_c)}{\lambda} \approx \ln(4e^\gamma a)$ holds under the condition $a = \frac{\omega_{\ln,D}}{2\pi T_c} = \frac{\omega_D}{2\pi T_c \sqrt{e}} \gg 1$. Likewise, $\frac{\mathcal{L}^D(T_c)}{\lambda}$ is approximately dominated by the logarithmic term if $a > 1$. However, the expression $T_c \propto \omega_{\ln}$ is not unique, as $\omega_{\ln}$ is also proportional to other average phonon frequencies. In this case, the definitive relation becomes $T_c \propto \omega_D$. To generalize the analysis, we assume that the acoustic branch of $\alpha^2 F(\omega)$ is fitted to the Debye model within the range $0 \leq \omega \leq \omega_D$. For $\omega_D < \omega \leq \omega_{\max}$, $\alpha^2 F(\omega)$ is mainly governed by optical phonons. By taking into account $\Omega \cong 2\pi T_c < \omega_{\ln,D}$, we derive a notable result: $\frac{\mathcal{L}(T_c)}{\lambda} \approx \left(\frac{0.040}{e}\right) \frac{\lambda_D}{\lambda} \frac{1}{a_D^2} + \ln(4e^\gamma a)$, which approximately ensures the relation $\frac{\mathcal{L}(T_c)}{\lambda} \approx \ln(4e^\gamma a)$ if $a_D > 1$. Consequently, we convey the necessity of $\omega_{\ln}$ through the analytic $T_c$ formula $T_c^{NTP} = \frac{2e^{\gamma-1}}{\pi} \omega_{\ln} \exp\left\{-\frac{1+\mu^*\ln(4e^\gamma m_c)}{\lambda}\right\}$. As a final remark, the mathematical procedures outlined in section 3 are anticipated to remain applicable to the disjoint type of $\alpha^2 F(\omega)$.

After the theoretical discussion, we illustrate the strong correlation between $T_c^{NTP}$ and $T_c^{AD}$, particularly with $T_c^{AD(\text{Mod})}$, in figure 1, where significant deviations from the unity line emerge when $T_c^{NTP} \gtrsim 120$ K. We suspect that this occurs due to the inclusion of $\omega_2$ in both $T_c^{AD}$ and $T_c^{AD(\text{Mod})}$, which is crucial in the high-$T_c$ regime, whereas $T_c^{NTP}$ does not account for it. In a rigorous manner, we prove that $a_D > 1$ is sufficient to guarantee the validity of $T_c^{NTP}$; on the other hand, $a > 1$ is a necessary condition. In the first line, the condition $a > 1$ may be practically adopted in preference to $a_D > 1$. As shown in figure 2, we compare $T_c^{NTP}$ with the exact numerical solutions to equation (9), where $m_c = 95$ is chosen. We notice that, on average, most data points are aligned along the unity line. Interestingly, a few data points are visibly located above the equality line, as marked by the green circle. Upon closer inspection, these data points do not meet the condition $a_D > 1$. In essence, although the condition $a > 1$ is satisfied, we observe that $a_D > 1$ may not be the case, as seen in the three data points in the top-right corner below the equality line.

**Appendix: Asymptotic behavior of $\frac{\Delta \mathcal{L}^E(a;m_c)}{\lambda}$ in the limit $m_c \gg 1$**

The Einstein model specifies the summation on the left-hand side of Eq. (9) in a form:

$$\frac{\mathcal{L}^E(a;m_c)}{\lambda} = a^2 \sum_{m=-m_c}^{m_c} \frac{1}{|2m-1|} \frac{1}{(m-1)^2 + a^2}, \tag{A1}$$

which can be rewritten as

$$1 + a^2 \sum_{n=1}^{m_c+1} \frac{1}{2n-1} \frac{1}{n^2 + a^2} + a^2 \sum_{n=1}^{m_c-1} \frac{1}{2n+1} \frac{1}{n^2 + a^2}. \tag{A2}$$

In the limit $m_c \gg 1$, Eq. (A2) simplifies to the approximate form given in Eq. (18). To justify this approximation, we compute the following difference (uncertainty):

$$\frac{\Delta \mathcal{L}^E(T_c; m_c)}{\lambda} \equiv \frac{\mathcal{L}^E(T_c; m_c + 1) - \mathcal{L}^E(T_c; m_c)}{\lambda}$$

$$= \frac{a^2}{2(m_c + 2) - 1} \cdot \frac{1}{(m_c + 2)^2 + a^2} + \frac{a^2}{2m_c + 1} \cdot \frac{1}{m_c^2 + a^2}. \tag{A3}$$

By taking $m_c \gg 1$, we can estimate that

$$\frac{\Delta \mathcal{L}^E(T_c; m_c)}{\lambda} \approx \frac{a^2}{(2m_c)(m_c^2 + a^2)} \left[ 1 + \frac{1}{1 + \frac{4m_c}{m_c^2 + a^2}} \right]. \tag{A4}$$

Therefore, $\frac{\Delta \mathcal{L}^E(T_c;m_c)}{\lambda a^2}$ rapidly tends to zero as $m_c \gg 1$, at a rate greater than $O\left(\frac{1}{m_c^3}\right)$. In addition, if $a \gg m_c$, Eq. (A4) reads

$$\frac{\Delta \mathcal{L}^E(T_c; m_c)}{\lambda a^2} \sim \frac{1}{m_c a^2} \ll O\left(\frac{1}{m_c^3}\right). \tag{A5}$$

**Acknowledgements**

Nattawut Natkunlaphat extends his heartfelt gratitude to his loving family, a great supporter, Ms. Aiyaluck Thitipipatchai, and an excellent mentor, Prof. Udomsilp Pinsook. In addition, the scholarship granted by the Graduate School of Chulalongkorn University in commemoration of the 72[nd] anniversary of His Majesty King Bhumibol Adulyadej is gratefully acknowledged.